\documentclass[aps,prl,twocolumn,groupedaddress]{revtex4}

\usepackage{graphicx}
\usepackage{dcolumn}
\usepackage{bm}

\begin{document}

\title{Zero differential resistance state  of two
dimensional electron systems in strong magnetic fields. }

\author{ A. A. Bykov }
\author {Jing-qiao Zhang}
\author{Sergey Vitkalov}
\email[Corresponding author: ]{vitkalov@sci.ccny.cuny.edu}
\affiliation{Physics Department, City College of the City University of New York, New York 10031, USA}
\author{A. K. Kalagin and A. K. Bakarov }
\affiliation{Institute of Semiconductor Physics, 630090 Novosibirsk, Russia}

\date{\today}

\begin{abstract}

Zero differential resistance state is found in response to direct current applied to 2D electron systems at strong magnetic field and low temperatures. Transition to the state is accompanied by sharp dip of negative differential resistance, which occurs above threshold value $I_{th}$ of the direct current. The state depends significantly on the temperature and is not observable above several Kelvins. Additional analysis shows lack of the linear stability of the 2D electron systems at $I>I_{th}$ and inhomogeneous, non-stationary pattern of the electric current in the zero differential resistance state.  We suggest that the dc bias induced redistribution of the 2D electrons in energy space is the dominant mechanism leading to the new electron state.

\end{abstract}

\pacs{}

\maketitle


The nonlinear properties of highly mobile two dimensional electrons
in AlGaAs/GaAs heterojunctions have attracted considerable attention recently. Strong oscillations of the longitudinal resistance induced by
microwave radiation have been found at magnetic fields which satisfy the condition  $\omega=n \times \omega_c$, where $\omega$ is the microwave frequency, $\omega_c$ is cyclotron frequency and $n$=1,2....\cite{zudov,engel}   At high levels of the microwave excitations the minima of the  oscillations can reach value close to zero.\cite{mani,zudov2,dorozh1,willett}  This so-called
zero resistance state (ZRS) has stimulated extensive theoretical interest.
\cite{andreev,durst,anderson,shi,vavilov,dmitriev}  At higher magnetic
field $\omega_c > \omega$ a considerable decrease of magnetoresistance
with microwave power is found \cite{engel,dorozh1,willett} which has
been  attributed to intra-Landau-level transitions.\cite{dorozh2}

Another interesting nonlinear phenomena have been observed in response to DC electric field.\cite{yang,bykov1,bykov2,zudov3}  Oscillations of the longitudinal resistance, periodic in inverse magnetic field, have been observed at relatively high DC biases satisfying the condition $n \times \hbar \omega_c=2 R_c E_H$, where $R_c$ is the cyclotron radius of electrons at the Fermi level and $E_H$ is the Hall electric field induced by the DC bias in the magnetic field. The effect has been attributed to "horizontal" Landau-Zener tunneling between Landau levels, tilted by the Hall electric field $E_H$.\cite{yang}. Another notable nonlinear effect: strong reduction of the longitudinal resistance by the dc electric field has been observed at considerably smaller biases \cite {bykov1,bykov2,zudov3}. In the paper\cite{bykov2} the strong reduction of the resistance  has been attributed to substantial changes of the electron distribution function induced by the DC electric field $E_{dc}$. Reasonable agreement has been established between the experiment and recent theory \cite{dmitriev}. 

In accordance with these findings\cite{dmitriev,bykov2} the nonlineariry is strongly enhanced at low temperature due to substantial decrease of the electron-electron scattering, which equilibrates the distribution function. A natural question arises: what happens at low temperatures, at which the nonlinearity is so strong, that the resistance drops very fast with the electric field? In this paper we investigate this question. We have found that at low temperatures the strongly decreasing differential resistance stabilizes suddenly near zero value at dc biases $I_{dc}$ above a threshold value $I_{th}$: $I_{dc}>I_{th}$. In other words, at $I_{dc}>I_{th}$ the longitudinal dc voltage $V_{xx}$ becomes independent on the dc bias.  The phenomenon is accompanied by sharp dip of the differential resistance at the threshold bias. At higher biases $I_{dc} > I_{th}$ temporal fluctuations of the differential resistance around the zero value are observed. Thus the experiments demonstrate that the dc biased 2D electron system in the strong magnetic field undergoes a transition to a quasi-stationary state with the zero differential resistance.   
At even higher biases, corresponding to the inter-Landau level transitions \cite{yang,bykov1,bykov2,zudov3,glazman} the differential resistance again becomes positive. 

Our samples are cleaved from a wafer of a high-mobility GaAs quantum well
grown by molecular beam epitaxy on semi-insulating (001) GaAs
substrates. The width of the GaAs quantum well is 13 nm. 
Three samples (N1,N2,N3) are studied with electron density $n_1$=8.2 $\times
10^{15}$ (m$^{-2}$), $n_2$=8.4$\times 10^{15}$ (m$^{-2}$), $n_3$=8.1$\times 10^{15}$ (m$^{-2}$)  and
mobility $\mu_1$=85 (m$^2$/Vs), $\mu_2$=70 (m$^2$/Vs) and $\mu_3$=82 (m$^2$/Vs) at T=2K.
Measurements are carried out between T=0.3K and T=20 K in magnetic
field up to 1 T  on $d$=50 $\mu m$ wide Hall bars with a distance of 250
$\mu m$ between potential contacts. The differential longitudinal resistance is
measured at a frequency of 77 Hz in the linear regime.  Direct electric current (dc bias) was applied simultaneously with ac excitation through the same current leads (see insert to fig. 1).  All samples demonstrate similar behavior. We show data for sample N1.

\begin{figure}
\includegraphics{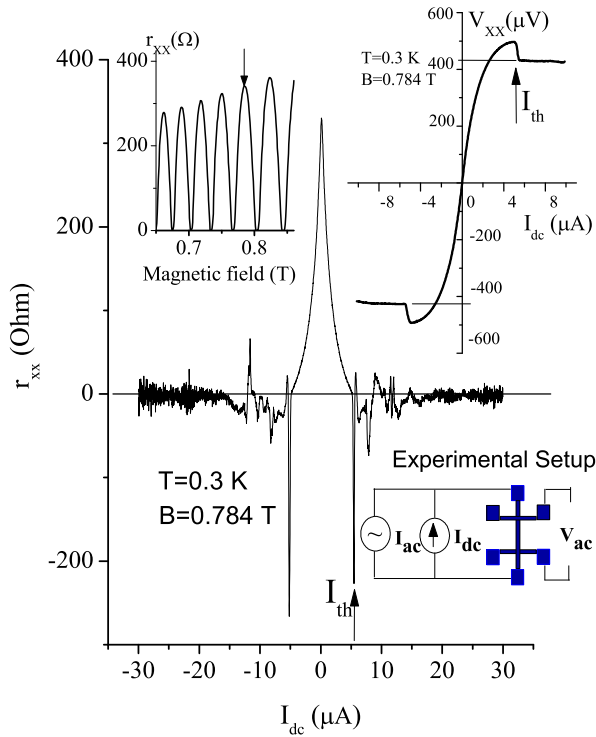}
\caption{ \label{fig1}  
Dependence of differential resistance on dc bias at T=0.3K and B=0.784T 
marked by arrow on top left insert.
Top left insert shows the linear resistance vs magnetic field at T=0.3K.
Top right insert shows $V_{xx}-I_{dc}$ dependence at T=0.3K and B=0.784T, threshold bias  $I_{th}$ is marked by arrow. Experimental setup is shown at bottom right.
  }
\end{figure}

Dependence of the longitudinal resistance $r_{xx}=dV_{xx}/dI$ on the DC bias is presented  in Fig. 1 at temperature T=0.3(K) and magnetic field B=0.784 (T). The magnetic field corresponds to one of maximums of Shubnikov de Haas (SdH) oscillations indicated by arrow in the top left insert to the figure. At small dc biases the differential resistance decreases approaching the zero. After touching the zero value the resistance demonstrates reproducible sharp dip at dc bias $I_{th}$=5.45 ($\mu$A) and, then the differential resistance $r_{xx}$ stabilizes near the zero value. At higher biases temporal fluctuations of the differential resistance (and/or the longitudinal voltage) are observed. The top right insert to the figure shows $V_{xx}$ vs $I_{dc}$ dependence corresponding to the same experimental conditions.  At dc biases below the threshold value the $V_{xx}-I_{dc}$ curve is essentially identical to one (not shown) obtained by the integration of the differential resistance. At dc biases above the threshold current $I_{th}$ a difference between these two curves is found. We believe that the difference is a result of the temporal fluctuations of the longitudinal voltage above the threshold $I_{th}$. It requires additional investigations and is beyond the scope of the paper. 

Fig.2a demonstrates the dc bias dependences of the $r_{xx}$ at different temperatures at fixed magnetic field B=0.784T. The transition to the zero differential resistance state is observed at temperatures below 4K. At higher temperatures the differential resistance, although becoming quite small, does not demonstrate the transition to the zero differential resistance state. A possible reason of such behavior is that at higher temperatures the nonlinear response to the dc bias is much weaker, due to substantial increase of the electron-electron scattering at high temperature\cite{dmitriev,bykov2}. The weaker nonlinearity requires stronger dc biases for the same nonlinear change of the resistance.  The higher dc bias makes the inter Landau level scattering to be more intensive \cite{yang,bykov1,bykov2,zudov3,glazman}, increasing the longitudinal resistance.  

 \begin{figure}
\includegraphics{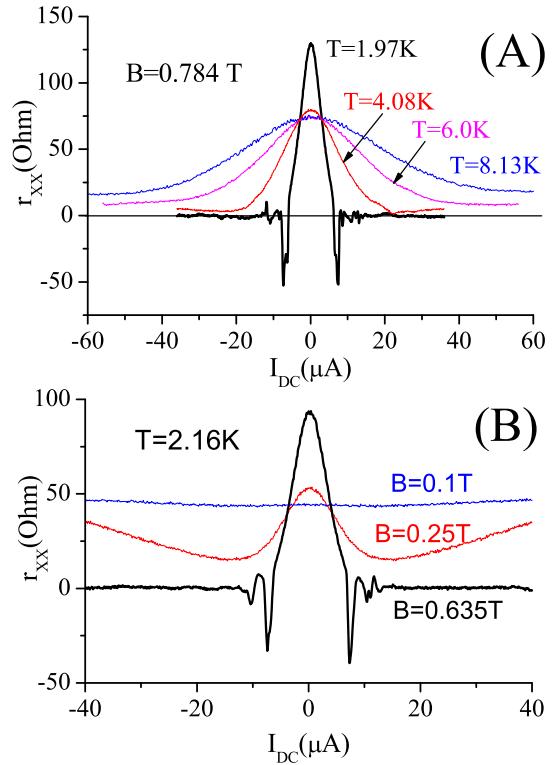}
\caption{ \label{fig2}  
a) Dependence of differential resistance $r_{xx}$ on dc bias 
at different temperature as labeled and B=0.784 T. The zero differential state 
is observed at T=1.97K
b) Dependence of differential resistance $r_{xx}$ on dc bias 
at different magnetic fields as labeled and  T=2.18K. The zero differential resistance state 
is observed at B=0.635T
}
\end{figure}

The density of electron states is highest at SdH maximums.  The high density of states enhances significantly the electron diffusion in the energy space providing strong nonlinear response to the dc bias\cite{dmitriev,bykov2}.  Fig.2b demonstrates the dependence of the $r_{xx}$ on the dc bias taken at different magnetic fields corresponding to SdH maximums. At T=2.16K the transition to zero differential resistance state occurs at magnetic fields above B=0.4 (T). At lower magnetic field the differential resistance reaches a minimum $r_{min}>0$ at a dc bias $I_{min}$.  The resistance $r_{min}$ increases with a decrease of the magnetic field\cite{bykov2}. The increase of the $r_{min}$ at small magnetic fields is due to, at least, two reasons. One is related to the decrease of the modulation of the density of states with the energy at low magnetic fields due to substantial overlap of  Landau levels. The weak modulation of the density of states makes the diffusion in the energy space to be more uniformed and, in result, the magnitude of the oscillating non-equilibrium distribution function and the nonlinear conductivity become to be small\cite{dmitriev,bykov2}. Another reason is the increase of the (bias induced) inter Landau level scattering due to the decrease of the Landau level separation at low magnetic field\cite{yang,glazman}.  Thus the transition to the state with the zero differential resistance occurs in 2D systems with substantial variations of the coefficient of the electron diffusion in the energy space, weak electron-electron relaxation and small inter-level scattering.

To the best of our knowledge  the zero differential resistance state is a new, strongly nonlinear state of the dc biased 2D electron systems in high magnetic field \cite{hallbreakdown}. However we believe that this state has strapping interrelations with the zero resistance state found in the microwave experiments\cite{mani,zudov2}. Our conclusion is based on a) the apparent relation between the zero differential resistance state and the strong decrease of the resistance due to the dc bias,   
b) experiments\cite{bykov2}, indicating that the bias induced variations of the distribution function \cite{dmitriev} is the dominant mechanism of the dc nonlinearity and c) convincing theoretical arguments demonstrating that the same nonlinear mechanism should be dominant both for the dc\cite{dmitriev} and the microwave\cite{dorozh1,dmitriev}  induced nonlinearities.

The important difference, however, between the zero resistance state (ZRS) and the zero differential resistance state is that the later does occur at relatively strong dc biases. The effect of the DC biases on the ZRS has been studied in paper \cite{mani2}. It was found that the microwave radiation makes the response to the dc electric field to be more linear. Strongly nonlinear I-V curves were observed in this experiment, however, the transition to the zero differential resistance state has not been detected (see also recent experiments \cite{bykov1,bykov2,zudov3}). We suggest that the $lower$ electron mobility in our samples can be the possible reason of the stronger nonlinearity. To support the suggestion we note, that in accordance with the theory\cite{dmitriev}  the nonlinear correction to the conductivity is inversely  proportional to the transport scattering time $\tau_{tr}$ and, therefore, can be stronger in lower mobility samples. The physical reason of such unusual behavior is the increase of the electron diffusion in the energy space due to the increase of the impurity scattering in low mobility samples.  This point is also supported by recent experimental observation of the microwave induced ZRS in the low mobility samples \cite{bykov3}. Thus very high mobility samples may not have the strongest nonlinear response.

Although the dominant mechanism of the nonlinearity in the two dimensional electron systems in strong magnetic fields becomes to be more apparent, the experimental study of the strongly nonlinear stationary states of 2D electrons require additional efforts. There are several theoretical proposals \cite{andreev,vavilov,auerbach,alicea,auerbach2}, which have analyzed the stability of non-equilibrium 2D electron systems in magnetic field. In this paper we use the approach developed in the article\cite{andreev}.  Assuming the local relation between electric field $\vec E$ and current density $\vec J$: $\vec E= \rho(J^2) \vec J$ and  taking into account continuity and Poisson equations the stability conditions has been found  \cite{andreev}:

\begin{figure}
\includegraphics{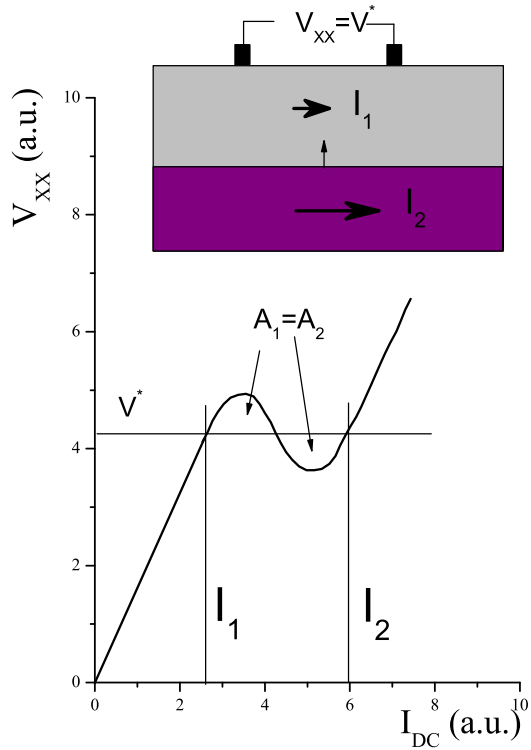}
\caption{ \label{fig3} 
The schematic picture of the of the current distribution 
for 2D electron system in  Hall bar geometry. 
The N-shaped $V_{xx}-I_{dc}$ dependence demonstrates two values $I_1$ and $I_2$ 
of the dc current corresponding to two homogeneous 
current distribution with the same longitudinal voltage $V^*$.  The boundary, which separates two regions with 
different currents,  moves from one side of the sample to 
another one with velocity determined by the equal-area 
rule: $A_1=A_2$\cite{butcher}.     
}
\end{figure}

$$
\rho_{xx}(\vec J^2) \ge 0 \eqno{(1)} 
$$
$$
\rho_{xx}(\vec J^2)+ \alpha \vec J^2 \ge 0 \eqno{(2)}  
$$

, where $\rho_{xx}$ is longitudinal resistivity and $\alpha=2(\frac{d\rho_{xx}(\vec J^2)}{d\vec J^2})$. 

In our case at the transition the longitudinal resistivity is positive $\rho_{xx}>0$. To analyze the second condition of the stability (eq.2) we note that $\rho_{xx}=E_x/J$ and $\frac{d\rho_{xx}}{dJ}=2J \frac{d\rho_{xx}}{d (J^2)}$, where $E_x$ is longitudinal component of the electric field $\vec E$. With these relations eq.2 can be rewritten in the following form:
$$
  \rho^{diff}_{xx}=dE_x/dJ \ge 0 \eqno{(2a)}
$$
,where $\rho^{diff}_{xx}$ is the longitudinal differential resistivity. The longitudinal differential resistance $r_{xx}$ differs from $\rho^{diff}_{xx}$ by a geometric factor $\gamma$ (in our samples $\gamma$=5): $r_{xx}=\gamma \rho^{diff}_{xx}$. Thus in accordance with eq.2 the 2D electron systems should be unstable at negative differential resistance: $r_{xx}<0$. Our experiment demonstrates transition to the zero resistance state at $r_{xx}=0$ in complete agreement with the theoretical consideration \cite{andreev}.  
We note also that the stability condition $r_{xx} \ge 0$ is quite general and has been used to analyze a broad set of instabilities, in particular Gunn effect\cite{bonch}. 

Using an analogy with the Gunn domain formation \cite{butcher,bonch}  the instability of the 2D homogeneous state at $r_{xx}<0$ leads to a domain structure with the electrical charge distribution (electric domain) moving from one boundary to another one. The domain velocity is determined by the I-V characteristic and may depend on the applied bias. A possible 2D current distribution has been considered recently\cite{vavilov}. In the proposed picture the electric current (and the Hall electric field) becomes to be bistable. In Hall bar geometry two regions with the stable currents $I_1$ and $I_2$ carry the same longitudinal voltage $V^{*}$ as shown in the fig.3. The boundary between these two regions must carry an additional electrical charge, because the Hall electric field changes across the boundary. Due to the presence of the longitudinal electric field $E^*_x=V^*/L$ (L is distance between potential contacts) the electrically charged boundary is not stationary and propagates across the sample with velocity: $v_y=c[E^*_x \times B]/B^2$, where $c$ is velocity of light.  Thus the electrical current $I(t)$ oscillates in time, with an averaged value equals to the dc bias: $<I(t)>=I_{dc}$. At the same time the longitudinal voltage $V_{xx}$ is independent on the dc bias   $I_{dc}$ and determined by the equal-areas rule ($A_1=A_2$ see fig.3) \cite{butcher}.  The independence of the longitudinal voltage $V_{xx}$ on the dc bias corresponds to our observation (see fig.1). 

In summary, zero differential resistance state $r_{xx}=0$ of the dc biased 2D electron systems is observed in strong magnetic fields at low temperatures. The transition to the state is accompanied by sharp dip  of the differential resistance and considerable temporal fluctuations of the longitudinal resistance arising above threshold value of the dc bias $I_{th}$.  We suggest that the quasi-stationary nonlinear electron state builds up due to local instability of the homogeneous distribution of the electric current at $r_{xx}<0$, which is  induced by significant, bias stimulated repopulation of electron states in the energy space.  The stability analysis of the 2D systems in magnetic fields supports strongly this suggestion.

\begin{acknowledgments}

We thank Prof. Myriam P. Sarachik for numerous discussions and technical
support. S. Vitkalov thanks Prof. I. Aleiner for discussion.  This work was supported by NSF: DMR 0349049 and RFBR, project
No.04-02-16789 and 06-02-16869. Department of Energy (DOE-FG02-84-ER45153) provided support for materials and equipment. 

\end{acknowledgments}


\begin{references} 

\bibitem{zudov} M.A. Zudov, R. R. Du, J. A. Simmons, and J. L. Reno, Phys. Rev. B {\bf 64},201311(R) (2001); \bibitem{engel}P.D. Ye, L. W. Engel, D.C. Tsui, J. A. Simmons, J. R. Wendt, G. A. Vawter, and J. L. Reno, Appl. Phys.Lett {\bf 79},2193 (2001).
\bibitem{mani} R. G. Mani, V.Narayanamurti, K. von Klitzing, J. H. Smet, W. B. Jonson, and V. Umansky, Nature(London) {\bf 420}, 646 (2002)
\bibitem{zudov2} M.A. Zudov, R. R. Du, L. N. Pfeiffer, and  K. W. West, Phys. Rev. Lett {\bf 90} 046807 (2003).
\bibitem{dorozh1} S. I. Dorozhkin, JETP Lett. {\bf 77}, 577 (2003).
\bibitem{willett} R. L. Willett, L. N. Pfeiffer, and  K. W. West, Phys. Rev. Lett {\bf 93} 026804 (2004). 
\bibitem{andreev} A. V. Andreev, I. L. Aleiner, and A. J. Millis, Phys.Rev. Lett. {\bf 91}, 056803 (2003)
\bibitem{durst} A. C. Durst, S. Sachdev, N. Read, and S. M. Girvin, Phys. Rev. Lett. {\bf 91}, 086803 (2003)
\bibitem{anderson} P. W. Anderson and W. F. Brinkman, cond-mat/0302129
\bibitem{shi} J. Shi and X. C. Xie, Phys. Rev. Lett. {\bf 91}, 086801 (2003).
\bibitem{vavilov} M. G. Vavilov and I. L. Aleiner Phys. Rev. B {\bf 69}, 035303 (2004)
\bibitem{dmitriev} I. A. Dmitriev, M.G. Vavilov, I. L. Aleiner, A. D. Mirlin, and D. G. Polyakov, Phys. Rev. B {\bf 71}, 115316 (2005).
\bibitem{dorozh2}S. I. Dorozhkin, J. H. Smet, V. Umansky and K. von Klitzing Phys. Rev. B {\bf 71}, 201306(R) (2005).
\bibitem{yang} C. L.Yang, J. Zhang, and R. R. Du, J. A. Simmons and J. L. Reno, Phys. Rev. Lett. {\bf 89}, 076801 (2002)
\bibitem{bykov1} A. A. Bykov, Jing-qiao Zhang, Sergey Vitkalov, A. K. Kalagin, and A. K. Bakarov   Phys. Rev. B {\bf 72}, 245307 (2005)
\bibitem{bykov2} Jing-qiao Zhang, Sergey Vitkalov, A. A. Bykov, A. K. Kalagin, and A. K. Bakarov   Phys. Rev. B {\bf 75}, 081305(R) (2007)
\bibitem{zudov3} W. Zhang, H.-S. Chiang, M. A. Zudov, L.N. Pfeiffer, and  K.W. West, Phys. Rev. B {\bf 75}, 041304(R) (2007) 
\bibitem{glazman} M.G. Vavilov, I.L Aleiner, and L.I. Glazman, cond-mat/0611130 
\bibitem{hallbreakdown} One should not directly relate the zero differencial resistance state  to the break down of the Quantum Hall Effect (see L.B. Rigal, et al. Phys. Rev. Lett. {\bf 82}, 1249 (1999)). The Quantum Hall Effect breaks down at the minimums of the quantum oscillations and accompanied by strong $increase$ of the longitudinal resistance. 
\bibitem{mani2} R. G. Mani, V. Narayanamurti, K. von Klitzing, J.H. Smet, W.B. Jonson, and V. Umansky Phys. Rev. B {\bf 70}, 155310 (2004)
\bibitem{bykov3} A. A. Bykov, A. K. Bakarov, D. R. Islamov, A. I. Toropov, JETP Letters {\bf 84}, 391 (2006)
\bibitem{auerbach} A. Auerbach, I Finkler, B. I. Halperin, and A. Yacoby, Phys. Rev. Lett. {\bf 94}, 196801 (2005)
\bibitem{alicea} J. Alicea, L. Balents, M.P.A. Fisher, A. Paramekanti, L. Radzihovsky, Phys. Rev. B {\bf 71}, 235322 (2005) 
\bibitem{auerbach2} Assa Auerbach, G. Venketeswara Pai cond-mat/0612469
\bibitem{bonch} V.L. Bonch-Bruevich, S.M. Kogan,  Sov. Phys. Solid-State {\bf 7}, 15 (1965)
\bibitem{butcher} P.N. Butcher, Phys. Lett. {\bf 19}, 546 (1965)




\end{references}

\end{document}